\documentclass[twoside,a4paper,12pt]{article}
\usepackage{color}  
\usepackage{amsfonts}                   
\usepackage[psamsfonts]{amssymb}
\usepackage[bottom]{footmisc}
\usepackage{amsmath}
\usepackage{multicol}
\usepackage{a4wide}                     
\usepackage{multicol}
\usepackage{lastpage}
\usepackage{geometry}
\def\ad{\mbox{ad}\,}
\def\b#1{{\mathbb #1}}                  
\def\ca#1{{\cal #1}}
\def\tfrac #1#2{\textstyle{\frac{#1}{#2}}}

\def\Dirac{{\raise0.09em\hbox{/}}\kern-0.69em D}

\newcommand{\initiate}{\setcounter{equation}{0}}

\begin{document}

\title{Kaluza-Klein Aspects of Noncommutative
  Geometry\thanks{Published in {\em Differential Geometric Methods in
  Theoretical Physics}, A.~I. Solomon, ed., pp.~243--252. World
  Scientific Publishing, 1989. Chester, August 1988.}}

\author{J. Madore 
\\[25pt]
Laboratoire de Physique Th\'eorique et Hautes Energies
\thanks{Laboratoire associ\'e au CNRS}
\\[10pt]
Universit\'e de Paris-Sud, B\^at. 210,  \ F-91405 ORSAY}
\date{}

\maketitle

\abstract{Using some elementary methods from noncommutative geometry a
structure is given to a point of space-time which is different from and simpler
than that which would come from extra dimensions. The structure is described by
a supplementary factor in the algebra which in noncommutative geometry replaces
the algebra of functions. Using different examples of algebras it is shown 
that the extra structure can be used to describe spin or isospin.}

\vfill

\initiate
\section{Introduction} 

Implicit in the formalism of a Kaluza-Klein version of
unified field theory is the hypothesis that what looks like a point in a
macroscopic description of space has in fact a richer structure at sufficiently
small length scales. The original attempt consisted in replacing a point by a
small circle by adding a fourth compactified dimension to the three we already
know. In this way it was possible to unify gravity and electromagnetism as a
theory of graviation in the resulting space-time of five dimensions. Subsequent
work replaced the circle by a more general manifold as a more complicated
structure was necessary to describe the extra gauge bosons which had been
discovered and to unify them with the gravitational field. We shall modify here
the original idea even more by supposing that the additional structure which a
point acquires at microscopic length scales is not that of a differential
manifold but is one which can only be described by generalizing the notion of a
manifold to include noncommutative geometry. The models we shall consider
however are in a sense which we shall make precise later even simpler than the
original model of Kaluza and Klein. The supplementary structure which we shall
use is algebraic in nature. It is too simple to give a phenomenologically
correct description of either the electro-weak or the strong interactions but
it allows us to change Maxwell's theory into an interesting gauge theory which
can be unified as usual with the gravitational theory to yield a unified theory
of Yang-Mills fields with gravity. 

We recall the definition of the manifold on which 
Kaluza-Klein theory is usually based. Locally it is of the form
\begin{align}
\b{R}^4 &\times F 
\nonumber\\ 
&\downarrow 
\nonumber\\ &\b{R}^4                                                           \label{1.1}
\end{align}
where $F$ is a manifold. We have therefore the imbedding
\begin{equation}
0 \rightarrow \ca{C} \rightarrow \ca{C} \otimes \ca{C}(F)       \label{1.2}
\end{equation} 
where $\ca{C}$ designates an algebra of complex-valued
functions, large enough to separate points. We have set 
$\ca{C}(\b{R}^4) = \ca{C}$ 
and we have supposed that the algebra of a product manifold can be 
identified with the product of the algebras of the individual factors.
Kaluza-Klein theory in the usual sense can be 
described equally well by referring to~(\ref{1.1}) or to~(\ref{1.2}). The internal 
structure is described by the manifold $F$ or by the 
algebra of functions $\ca{C}(F)$.

We can now generalize Kaluza-Klein theory by replacing $\ca{C}(F)$ by
an associative algebra which is not necessarily commutative. We can
then no longer refer to the diagram~(\ref{1.1}) as there is no
internal manifold $F$.  We shall choose as algebra one of the matrix
algebras $M_n$ of complex $n \times n$ matrices. These choices have
several advantages the most important being related to the fact that
the space of derivations of a matrix algebra is of finite dimension.
It follows in particular that the internal laplacian has a finite
spectrum and that the total momentum space remains of
dimension~4. This is important for the renormalizability of the
theory.

The noncommutative factor is the origin of the 
extra structure which a point acquires.
The elements of the resulting algebra $\ca{A}$ are what replace the
functions on $\b{R}^4$. In particular the nearest object which we have to a
coordinate is an element of this algebra. This means that the position of a
particle, for example, no longer has a well-defined meaning. Since we
certainly wish this to be so at macroscopic scales, we must 
require that the scale $\kappa$ at which $\ca{A}$ differs from
$\ca{C}$ be not 
much greater than a typical Compton wave length. In other words, the 
fuzziness which the internal structure gives a point in space-time cannot 
be much greater than the quantum uncertainty in the position of a particle.
We shall suppose that the 
Lorentz group acts on $\ca{A}$. This means that
directions are well-defined. Our space-time looks like a crystal then which
has a homogeneous distribution of dislocations but no disclinations. We can
pursue this solid-state analogy and think of the ordinary Minkowski
coordinates as macroscopic order parameters obtained by course-graining over
scales less than $\kappa$. They break down and must be replaced by elements
of the algebra $\ca{A}$ when one considers phenomena on these scales.
The models which we propose can be considered as a 
classical phenomenological description of an additional microscopic 
structure. They do not rule out the 
possibility that the different mass scales which one observes in nature are 
of quantum origin since we do not rule out the possibility that the 
extra structure which we shall use is due to quantum effects.

The extra structure which a point acquires at length scales less than
the length scale $\kappa$ could be considered as the origin of the
spin of a particle. This possibility is explored in Section~3. See
also [3,~7].  It could also be identified with the origin of
isospin. This possibility is explored in Section~4.  See also
[4,~5,~6]. In Section~5. we describe briefly the noncommutative
version of electromagnatism as an application of the formalism
introduced in Section~4.  We refer to [1] for a general introduction
to noncommutative geometry and for references to the previous
literature.  We refer to [2,~3,~4,~5,~6] for more details on the
particular models which we shall consider here.

\initiate
\section{Mathematical Preliminaries}

Differential geometry was developed to study the structure of manifolds and of
the algebras of functions which are defined on them. Noncommutative geometry
was invented to extend this study to associated algebras in general. The matrix
algebras we shall use are simple and there is no need to introduce geometric
methods to study them except perhaps as examples. The geometry of matrices
has however as we shall see an interesting application in describing the extra
physical structure we mentioned in the preceeding section. 

A manifold is completely determined by the algebra of
functions defined on it. All of the geometrical quantities one uses to describe
classical physics such as vector fields, differential forms, metrics and
connections can be reformulated as operations involving only this algebra. 
One can therefore develop a noncommutative version of classical physics
simply by replacing the quantities in the commutative case which are used to
describe the theory by the corresponding operations on a general 
associative algebra $\ca{A}$. For example
a scalar field becomes an element of an abstract algebra which is no longer an
algebra of functions. The scalar field then is no longer a function on
space-time. 
If $\ca{A}$  is commutative then under certain very general conditions, it
can be shown to be an algebra of functions. If $\ca{A}$  is too abstract or
too different from the ordinary algebra of functions on space-time, it is
impossible to interpret its elements in terms of classical observables. As a
compromise between these two extremes, and for the 
reasons which were given above,
we shall use here as algebra the tensor product of the algebra $\ca{C}$ of 
smooth complex-valued functions on space-time and the algebra $M_n$
of complex $n \times n$ matrices: 
\begin{equation}
\ca{A} = \ca{C}\otimes M_n .
\end{equation}

The Lie algebra $D(\ca{C})$ of smooth vector fields on the manifold
$\b{R}^4$ can be identified with the algebra of derivations of
$\ca{C}$, that is, with the algebra of linear maps of $\ca{C}$ into
itself which satisfy the Leibnitz rule. This algebra is the most
important mathematical object which one uses when one studies
classical fields on $\b{R}^4$ and their dynamics. To study these
fields then in the noncommutative case we must consider the
derivations $D(\ca{A})$ of the algebra $\ca{A}$. This Lie algebra is the
direct sum of the ordinary derivations of $\ca{C}$ and the
$\ca{C}$-module generated by the inner derivations of $M_n$. A problem
which arises which was not present in the commutative case is the fact
that the derivations no longer form a module over the algebra
$\ca{A}$. A derivation when multiplied by an element of $\ca{A}$ is no
longer a derivation. This is in sharp contrast to the normal situation
where a vector field multiplied by a function remains a vector
field. We avoid this problem as much as possible by working with the
generalizations of differential forms, which are dual to the
derivations.

Let $\lambda_a$, for $1 \leq a \leq n^2-1$,
be a basis of the Lie algebra of the 
special unitary group in $n$ dimensions, chosen so that the structure 
constants $C^a{}_{bc}$ are real. The Killing metric is 
given by $g_{ab} = -Tr(\lambda_a \lambda_b)$. 
We shall raise and lower indices with this metric.
The set $\lambda_a$ is a set of generators of the matrix algebra $M_n$.
It is not a minimal set but it is
convenient because of the fact that
the derivations $e_a = \kappa^{-1} ad (\lambda_a)$ form a basis 
over the complex numbers of the derivations of $M_n$. 
They satisfy the commutation relations
\begin{equation}
[e_a, e_b] = m \, C^c{}_{ab} \, e_c.                            \label{2.1}
\end{equation} 
The mass scale $m$ is defined to be the inverse 
of the length scale $\kappa$. 
Let $x^{\mu}$ be coordinates of $\b{R}^4$. Then
the set $(x^{\mu}, \lambda^a)$ is a set of generators of the algebra 
$\ca{A}$. We define the exterior derivative of an element of $\ca{A}$
as usual. For example, if $f$ is an element of $M_n$, then $df$ 
is defined [2] by the formula
\begin{equation}
df(e_a) = e_a(f).                                                       \label{2.2}
\end{equation}
Because of the particular structure of $M_n$ [4] there is a
system of generators of $\Omega^1 (M_n )$, the 1-forms on $M_n$,
completely characterized by the equations 
\begin{equation}
  \theta^a(e_b) = \delta^a_b.                                   \label{2.3}
\end{equation}
It is related to $d\lambda^a$ by the equations
\begin{equation}
d\lambda^a = m \, C^a{}_{bc}\, \lambda^b \theta^c,  \qquad
\theta^a = \kappa\,\lambda_b \lambda^a d\lambda^b.   \label{2.4}
\end{equation}

From the $\theta^a$ we can construct a 1-form $\theta$ 
in $\Omega^1 (M_n )$,
\begin{equation}
\theta = - m \lambda_a \theta^a,                                                \label{2.5}
\end{equation}
which from~(\ref{2.4}) satisfies the zero-curvature condition:
\begin{equation}
d \theta + \theta^2 = 0.                                                                        \label{2.6}
\end{equation}
We shall see below in Section 3 that $\theta$ is gauge invariant. It
satisfies with respect to the algebraic exterior derivative~(\ref{2.2}) similar
conditions to those which the Maurer-Cartan form satisfies with respect to
ordinary exterior derivation on the group $SU_n$. 
Choose a basis $\theta^{\alpha}_{\lambda} dx^{\lambda}$
of $\Omega^1(\ca{C})$ over $\ca{C}$ and let $e_{\alpha}$ be the 
pfaffian derivations dual to $\theta^{\alpha}$.
Set $i = (\alpha, a), \; 1 \leq i \leq  4 + n^2 - 1$, and
introduce $\theta^i = (\theta^{\alpha}, \theta^a )$ as generators of 
$\Omega^1 (\ca{A})$ as a left or right $\ca{A}$-module and 
$e_i = (e_{\alpha}, e_a)$ as a basis of $D(\ca{A})$ over $\ca{C}$. 

We shall introduce the quadratic form, of signature $d-2$, given by
\begin{equation}
ds^2 = \eta_{ij} \theta^i \otimes \theta^j =
\eta_{\alpha\beta} \theta^{\alpha} \otimes \theta^{\beta} +
g_{ab} \theta^a \otimes \theta^b.                                       \label{2.7}
\end{equation}
The $\eta_{\alpha \beta}$ is the Minkowski metric. We shall refer to this
quadratic form as a metric although it contains two terms of a slightly
different nature.

\initiate
\section{Spin} 

Consider $\b{R}^3$ with euclidean coordinates $x^a$. 
To the $x^a$ we associate  
operators $q^a$, self-adjoint elements of an algebra $\ca{A}$
which do not necessarily commute: 
\begin{equation}
q^{a*} = q^a, \qquad    q^a q^b \not=  q^b q^a,
\end{equation}
such that $\ca{A}$ is generated by them. We shall suppose that the  
euclidean  group acts on $\ca{A}$ by automorphisms induced by  
\begin{equation}
q^a \mapsto R^a_b q^b + a^b.
\end{equation}
It follows that the commutator $[q^a, q^b]$ is invariant under space  
translations. We suppose further that each $q^a$ has a unique decomposition
\begin{equation}
q^a = x^a + \kappa \sigma^a                                                             \label{3.1}
\end{equation}
as the sum of two elements the first of which belongs to the  
center of $\ca{A}$ and the second is invariant under the action of  
the translations. When the length scale $\kappa$ tends to zero the  
$q^a$ tend to the ordinary euclidean coordinates we started with.  
If we choose $\sigma^a$ to be the Pauli matrices
the algebra $\ca{A}$ splits as a tensor product of an algebra of functions
times the algebra of $2 \times 2$ matrices $M_2$. 
The extra factor is in some respects like an internal space  
in a Kaluza-Klein theory. 

There is an obvious generalization of the Ansatz~(\ref{3.1}) to a  
Poincar\'e invariant algebra with the generators 
\begin{equation}
q^{\alpha} = x^{\alpha} + \kappa \gamma^{\alpha}.                               \label{3.2}
\end{equation} 
The Dirac matrices $\gamma^{\alpha}$ are however not self-adjoint 
and the above  
defined $q^{\alpha}$ cannot all have real eigenvalues. We are naturally  
led to introduce an operator-valued Dirac spinor $z$ and consider  
the algebra $\ca{A}$ generated by  
\begin{equation}
q^{\alpha} = x^{\alpha} + \kappa J^{\alpha}, \qquad
J^{\alpha}= \bar z \gamma^{\alpha} z.                                                   \label{3.3}
\end{equation} 
We shall impose on $z$ the following commutation relations 
\begin{equation}
[z,z] = 0, \quad [z, \bar z] = 1, \quad   [\bar z, \bar z] = 0.
\end{equation}
The generators $q^{\alpha}$ do not commute then and we have
\begin{equation}
[q^{\alpha}, q^{\beta}] = - 2i \kappa^2  S^{\alpha\beta}                        \label{3.4}
\end{equation}   
where we have set    
\begin{equation}
S^{\alpha\beta} = \bar z \sigma ^{\alpha\beta} z,\qquad  
\sigma ^{\alpha\beta}  = \tfrac 12 i[\gamma^{\alpha}, \gamma^{\beta}].
\end{equation}    
The elements $(\gamma^{\alpha},\sigma ^{\alpha\beta})$ 
of the Clifford algebra generate the  
Lie algebra of $SO(3,2)$. So $SO(3,2)$ acts on $\ca{A}$. The conformal  
group would act also on the extension $\ca{A}^c$ of $\ca{A}$ 
obtained by adding  
the generator $\bar z\gamma^5 z$. The set of derivations
$D(\ca{A})$ of $\ca{A}$ is a  
module over the center of $\ca{A}$ generated by    
\begin{equation}
(\partial_{\alpha},\; \tfrac 12 i\ad \, J^{\alpha},\;
\tfrac 12 i\ad \, S^{\alpha\beta}).
\end{equation} 
Although the $(J^{\alpha}, S^{\alpha\beta})$ 
satisfy the commutation relations of  
the Lie algebra of $SO(3,2)$, they generate in fact an infinite-dimensional  
algebra as do the $x^{\alpha}$ . This means there is an internal structure  
more complicated than that which followed from the 
non-relativistic Ansatz given above and also from the Ansatz used in the
following sections.

Define the algebras $\ca{I}$ and $\ca{J}$ generated respectively 
by $(z, \bar z)$  
and by $(J^{\alpha}, S^{\alpha\beta})$. 
We are interested in $\ca{J}$ which is a subalgebra  
of $\ca{A}$, and of $\ca{I}$. We start then by  
studying $\ca{I}$ which we can consider as the quantized version  
of an algebra of functions over the classical phase space $(z,\bar z)$ 
with bracket
\begin{equation}
\{ z, \bar z \} = i.                                                            \label{3.5}
\end{equation}   
There are therefore in principle two distinct quantization procedures,
the ordinary one involving $\hbar$ and this new one. 
Under complete dequantization $\ca{A}$ becomes a commutative algebra  
with a Poisson bracket on $\ca{J}$ induced by~(\ref{3.5}).
To the $(x^{\alpha}, z, \bar z)$ given above we add $p_{\alpha}$ to form  
a phase space and we extend the bracket~(\ref{3.5}). The resulting phase space 
is related to one which has been studied in detail by Souriau~[8] and 
which can be conveniently used to describe a classical spinning particle.

\initiate
\section{Isospin}

In the commutative case a connection $\omega$ on the trivial principal 
$U_1$-bundle equipped with the associated canonical flat connection
is an anti-hermitian 1-form which can be split as the sum of a 
horizontal part, a 1-form on the base manifold, and a
vertical part, the Maurer-Cartan form $d\alpha$ on $U_1$, 
\begin{equation}
\omega = A + d\alpha.                                                                   \label{4.1}
\end{equation}
The gauge potential $A$ is an element of $\Omega^1(\ca{C})$ and using it we
can construct a covariant derivative on an associated vector
bundle. The notion of a vector bundle can be generalized to the noncommutative
case as an $\ca{A}$-module which in its simplest form, a free module of rank
1, can be identified with $\ca{A}$ itself. This is in fact the natural
generalization to the algebra we are considering of a trivial $U_1$-bundle
since $M_n$ has replaced $\b{C}$ in our models. So the $U_n$ gauge symmetry
we shall use below comes not from the rank of the vector bundle, which we shall
always choose to be equal to 1, but rather from the factor $M_n$ in our algebra
$\ca{A}$. The noncommutative generalization of $A$ is an anti-hermitian
element of
$\Omega^1(\ca{A})$, which in turn can be split
as the sum of two parts, called also horizontal and vertical. We shall here
designate by $\omega$ such an element of $\Omega^1(\ca{A})$ since we 
wish to reserve the letter $A$ and the name gauge potential
for the horizontal part in this latter sense. We write then 
\begin{equation}
\omega = A + \theta + \phi,                                                     \label{4.2}
\end{equation}
where $A$ is an element of $\Omega^1_H$ and $\phi$ is an element of
$\Omega^1_V$. The field $\phi$ is the Higgs field.  We have here
separated out the 1-form $\theta$ which is in many respects like a
Maurer-Cartan form.  Formula~(\ref{4.2}) with $\phi = 0$ and
formula~(ref{4.1}) are formally similar but the meaning of the words
horizontal and vertical in the two cases is not the same.

Let $\ca{G}$ be the group of invertible elements of $\ca{A}$, considered
as functions on $\b{R}^4$ with values in $GL_n$ and $g$ an element of
$\ca{G}$. 
Let $\ca{U}_n$ be the subgroup of $\ca{G}$ of elements which satisfy 
$gg^* = 1$. We shall choose it to be
the group of local gauge transformations. 

A gauge transformation defines a mapping of 
$\Omega^1(\ca{A})$ into itself of the form
\begin{equation}
\omega^{\prime} = 
g^{-1} \omega g + g^{-1} dg.                                                    \label{4.3}
\end{equation}
We require that $\phi$ transform under the adjoint action 
of $\ca{U}_n$: 
\begin{equation}
\phi^{\prime} = g^{-1} \phi g.                                                          \label{4.4}
\end{equation}
It can be seen then that $\theta$ is invariant under $\ca{G}$, 
\begin{equation}
\theta^{\prime} = \theta,                                                               \label{4.5}
\end{equation}
and so the transformed potential $\omega^{\prime}$ is again of the form~(\ref{4.2}).

We define the curvature 2-form $\Omega$ and the field strength $F$ as usual.
In terms of components, with
$\phi = \phi_a \theta^a$ and $A = A_{\alpha} \theta^{\alpha}$ and with
\begin{equation}
\Omega = \tfrac 12 \Omega_{ij} \theta^i \wedge \theta^j, \quad
F = \tfrac 12 F_{\alpha \beta} \theta^{\alpha} \wedge \theta^{\beta},   \label{4.6}
\end{equation}
we find
\begin{equation}
\Omega_{\alpha \beta} = F_{\alpha \beta}, \quad   
\Omega_{\alpha a} = D_{\alpha} \phi_a ,   \quad
\Omega_{ab} = [\phi_a,\phi_b] - m\,C^c{}_{ab}\,\phi_c.                  \label{4.7}
\end{equation}

We wish to use $\omega$
and a linear connection on $\Omega^1(\ca{C})$ to construct a 
linear connection on $\Omega^1(\ca{A})$. To simplify the calculations
we shall here suppose also that
\begin{equation}
Tr(\omega_a ) = 0,  \qquad   Tr(A_{\alpha} ) = 0,                                       \label{4.8}
\end{equation}
and that $g \in \ca{SU}_n$, the local $SU_n$ gauge transformations. 

First we restrict our considerations to that special class 
of connections whose vertical component $\omega_V$, 
in $\Omega^1_V$, is equal to the canonical 1-form $\theta$:
\begin{equation}
\omega_V = \theta.                                                                              \label{4.9}
\end{equation}

The curvature of the connection $\omega$ is constructed using the exterior 
derivative $d$ which acts on the coefficients $\omega^a$ as well as on the 
basis $\lambda_a$. We wish to introduce an effective exterior derivative 
which acts only on $\omega^a$ but which includes the action of $d$ on 
$\lambda_a$. Set
\begin{equation}
(\tilde d\omega^a)\lambda_a = d(\omega^a \lambda_a).                    \label{4.10}
\end{equation}
This can be rewritten as
\begin{equation}
\tilde d \omega^a = 
d \omega^a + m \, C^a{}_{bc}\, \omega^b \wedge \theta^c.                         \label{4.11}
\end{equation}
Define $\tilde \theta^a$ by the equation
\begin{equation}
\omega = -m \lambda_a\tilde\theta^a
\end{equation}
and set
\begin{equation}
\tilde \theta^i = (\theta^{\alpha}, \tilde \theta^a).
\end{equation}

Let $\omega^{\alpha}{}_{\beta}$ now be a linear connection on 
$\Omega^1(\ca{C})$, an $so(3,1)$-valued 1-form satisfying the 
structure equations
\begin{align}
&d\theta^{\alpha} + 
\omega^{\alpha}{}_{\beta} \wedge \theta^{\beta} = 0, \\
&d\omega^{\alpha}{}_{\beta} +
\omega^{\alpha}{}_{\gamma} \wedge \omega^{\gamma}{}_{\beta} =
\Omega^{\alpha}{}_{\beta}.                                                                       \label{4.12}
\end{align}

We must construct an $so(n^2 + 2,1)$-valued 1-form $\tilde \omega^i{}_j$
on $\Omega^1(\ca{A})$ satisfying 
the first structure equation  
\begin{align}
\tilde d\tilde \theta^i 
+ \tilde \omega^i{}_j \wedge \tilde \theta^j = 0.                       \label{4.13}
\end{align}
This construction is as usual.
One can see that the equations~(\ref{4.13}) are satisfied if 
$\tilde \omega^i{}_j$ 
is given by the equations
\begin{align}
&\tilde\omega^{\alpha}{}_{\beta} = \omega^{\alpha}{}_{\beta} -
\tfrac 12 \kappa^2 F_a{}^{\alpha}{}_{\beta} \, \omega^a, 
\nonumber\\
&\tilde\omega^a{}_b =
-\tfrac 12 C^a{}_{bc} \, \omega^c, 
\nonumber\\
&\tilde\omega^a{}_{\alpha} =
-\tfrac 12 \kappa F^a{}_{\alpha \beta} \, \theta^{\beta}, 
\nonumber\\
&\tilde\omega^{\alpha}{}_a =
\tfrac 12 \kappa F_a{}^{\alpha}{}_{\beta} \, \theta^{\beta}. 
\nonumber\\                                                             \label{4.14}
\end{align}

Consider next a general $\ca{SU}_n$ connection. Equation~(\ref{4.13}) becomes
\begin{equation}
\tilde d\tilde \theta^i + \tilde \omega^i{}_j \wedge \tilde \theta^j = 
\tilde\Theta^i.                                                                         \label{4.15}
\end{equation}
with a non-vanishing torsion form given by
\begin{equation}
\tilde\Theta^{\alpha} = 0, \qquad \lambda_a\tilde\Theta^a =
-\kappa (D\phi - \phi^2).                                                       \label{4.16}
\end{equation}

The last step in the Kaluza-Klein 
construction is to consider the second structure equations
\begin{equation}
\tilde d\tilde \omega^i{}_j + \tilde \omega^i{}_k \wedge \tilde \omega^k{}_j =
\tilde \Omega^i{}_j,                                                            \label{4.17}
\end{equation}
and the equations of motion which follow from a suitable action.
We shall not discuss this here.

\initiate
\section{ A Model}

 As action it is natural to choose the most general 
expression, polynomial in the Riemann tensor, which would 
yield second-order field equations. This yields in the absence of gravitation
the bosonic action
\begin{equation}
S = \tfrac 14 Tr \int(\Omega_{ij}\Omega^{ij}). 			\label{5.1}
\end{equation}
The integration is over space-time and the trace is what replaces this 
integration on the factor $M_n$ of $\ca{A}$. This is the simplest gauge
action which can be written down in the geometries which we consider.
We have put the Yang-Mills coupling constant equal to one.
Written out explicitly the lagrangian becomes
\begin{equation}
\ca{L} = \tfrac 14 Tr(F_{\alpha\beta} F^{\alpha\beta}) 
+ \tfrac 12 Tr(D_{\alpha}\phi_a D^{\alpha}\phi^a) - V(\phi),   			\label{5.2}
\end{equation}
where the potential $V(\phi )$ is given by
\begin{equation}
V(\phi) = -\tfrac 14 Tr(\Omega_{ab} \Omega^{ab}).   			\label{5.3}
\end{equation}
It is a quartic polynomial in $\phi$ which is fixed 
and has no free parameters apart from the mass scale $m$.

From~(\ref{4.7}) we see that $V(\phi )$ vanishes for the values
\begin{equation}
\phi_a = 0, \qquad \phi_a = m\,\lambda_a.     			\label{5.4}
\end{equation} 
There are therefore 2 stable phases. 
We shall consider first the symmetric phase, 
$\phi_a = 0$ and then the broken phase, $\phi_a = m\,\lambda_a$.
In the symmetric 
phase the masses of all the scalar modes are equal and they are real since 
the corresponding value of the potential is a stable minimum. 
Using the expression for $\Omega_{ab}$,
we find that the mass is given by
\begin{equation}
m^2_H = n m^2.  			\label{5.5}
\end{equation}
The gauge bosons of course have vanishing mass in the symmetric phase.
In the broken phase, the masses of the gauge bosons 
is given therefore by the equation
\begin{equation}
m^2_A = 2n m^2.   			\label{5.6}
\end{equation}

There is a certain ambiguity in the definition of a spinor
and the associated Weyl operator. It is possible to treat the basis~(\ref{2.3})
of $\Omega^1 (M_n )$ as though it were a moving frame on a manifold of
dimension $n^2 -1$. Using the metric 
which we introduced in Section 2, one can then construct
spinors and Dirac matrices in 
dimension $4 + n^2 - 1$ and introduce a Weyl operator in exactly the 
same way which one does in dimension $4$. We shall suppose here that this 
has been done.
Using the gauge lagrangian from the  previous section we can write the
lagrangian for noncommutative electrodynamics as
\begin{equation}
\ca{L} = \tfrac 14 Tr(\Omega_{ij}\Omega^{ij}) +
Tr(\bar \psi \Dirac \psi).  			\label{5.7}
\end{equation}
The interaction of the spinors with the gauge bosons is given by the term
\begin{equation}
\ca{L}_I = Tr ( \bar \psi \gamma^{\alpha} A_{\alpha} \psi). 			\label{5.8}
\end{equation}
So we see then that in the broken phase the zero-mass gauge 
boson is coupled to all
fermions including the one which has zero mass.

The above lagrangian~(\ref{5.7}) is a generalization of the Yang-Mills-Higgs-Kibble
lagrangian, with a more elaborate Higgs sector. Since we have replaced 
complex-valued functions by functions with values in the matrix algebra
$M_n$ it is to be expected that each $\phi_a$ takes its values
in $M_n$. The most original part is the potental term
$V(\phi )$ which comes from the curvature of the vertical part of the
connection. 
Consider the case $n=2$ and compare the above model with that of the
standard model for the electroweak interactions.
There is no analogue of the Weinberg angle. The massive gauge
bosons are neutral and they have all equal masses.
In the case $n=3$ we would have a model for gluons plus a spurious $U_1$
abelian gauge potential provided we suppose that we are in the symmetric 
phase. The major difference with the standard model for the strong interactions
lies in the fact that here the symmetric phase is a stable phase. 
The Higgs bosons are coloured and would not appear as asymptotic states.

{\it This work was done in collaboration with 
Michel Dubois-Violette and Richard Kerner.}


\begin{thebibliography}{10}


\bibitem {[1]}
J. Dixmier, Les C*-alg\`ebres et leurs repr\'esentations,  
Gauthier-Villard (1964), 
A. Connes, Publications of the I.H.E.S., {\bf 62} (1986) 257.

\bibitem {[2]} M. Dubois-Violette, 
C. R. Acad. Sci. Paris, {\bf 307}, S\'erie I (1988) 403.

\bibitem {[3]} J. Madore, "Non-commutative Geometry and the Spinning Particle",
Lecture given at the Journ\'ees Relativisites, 
Geneva, April 1988 and at the XI Warsaw Symposium on Elementary Particle 
Physics, Kazimierz, May 1988.

\bibitem {[4]} 
M. Dubois-Violette, R. Kerner, J. Madore, "Non-Commutative Differential
Geometry of Matrix Algebras", Orsay preprint, October 1988 (to appear). 

\bibitem {[5]} 
M. Dubois-Violette, R. Kerner, J. Madore, "Classical Bosons in a 
Noncommutative Geometry" Orsay preprint, November 1988 (to appear). 

\bibitem {[6]}
M. Dubois-Violette, R. Kerner, J. Madore, "Non-Commutative Differential
Geometry and New Models of Gauge Theory" Orsay preprint, 
November 1988 (to appear). 

\bibitem {[7]} H. Bacry, "The Notions of Localizability and Space: From Eugene
Wigner to Alain Connes", Talk given at the International Symposium on Space-
Time Symmetries, Univ. of Maryland, May 1988.

\bibitem {[8]} J.-M. Souriau, "Structures des Syst\`emes Dynamiques", Dunod,  
Paris (1969). See also C. Duval, "The Spin Polarizer" in Diff.  
Geom. and Math. Phys., M. Cahen et al. eds. Reidel (1983);   


\end{thebibliography}

\providecommand{\href}[2]{#2}\begingroup\raggedright\endgroup

\end{document}